\documentstyle[prb,twocolumn,aps,graphicx]{revtex}
\begin{document}
\draft

 \newcommand{\mytitle}[1]{
 \twocolumn[\hsize\textwidth\columnwidth\hsize
 \csname@twocolumnfalse\endcsname #1 \vspace{1mm}]}

\mytitle{
\title{Theory of Magnetic Properties and Spin-Wave Dispersion for
        Ferromagnetic (Ga,Mn)As}
\author{J\"urgen K\"onig,$^{1,2,3}$ T. Jungwirth,$^{2,3,4}$ and 
        A. H. MacDonald$^{2,3}$}
\address{$^1$Institut f\"ur Theoretische Festk\"orperphysik, 
        Universit\"at Karlsruhe, D-76128 Karlsruhe, Germany\\
        $^2$Department of Physics, University of Texas, Austin, TX 78712, USA
        \\
        $^3$Department of Physics, Indiana University, Bloomington, IN 47405, 
        USA
        \\
        $^4$Institute of Physics ASCR, Cukrovarnick\'a 10, 162 00 Praha 6, 
        Czech Republic}
\date{\today}
\maketitle

\begin{abstract}

We present a microscopic theory of the long-wavelength magnetic properties of
the ferromagnetic diluted magnetic semiconductor (Ga,Mn)As.
Details of the host semiconductor band structure, described by a six-band 
Kohn-Luttinger Hamiltonian, are taken into account.
We relate our quantum-mechanical calculation to the classical micromagnetic 
energy functional and determine anisotropy energies and exchange constants.
We find that the exchange constant is substantially enhanced compared to the 
case of a parabolic heavy-hole-band model.

\end{abstract}
}
%
%

\section{Introduction}

The recent discovery of carrier-mediated ferromagnetism in diluted magnetic 
semiconductors (DMS) has generated intense interest, in part because it 
suggests the prospect of developing devices which combine information 
processing and storage functionalities in one material.%
\cite{Prinz95,Prinz98,Furdyna,Dietl94,Pashitski,Story86,Haury97,Ohno92,Ohno96,%
Ohno98,Ohno99,Hayashi98,Pekarek98,VanEsch97/1,VanEsch97/2,Oiwa99,Matsukura98,%
Ohno98.2,Okabayashi98,Omiya00,Beschoten99}
Ferromagnetism has been observed in Mn doped GaAs up to critical 
temperatures $T_c$ of $110 {\rm K}$ (see Ref.~\onlinecite{Ohno99}).
Doping a III-V compound semiconductor with Mn introduces both local magnetic 
moments, with concentration $N_{\rm Mn}$ and spin $S=5/2$, and itinerant 
valence-band carriers with density $p$ and spin $s=1/2$.
An antiferromagnetic interaction between both kinds of spin mediates an 
effective ferromagnetic interaction between ${\rm Mn}^{2+}$ spins.

A phenomenological long-wavelength description of ferromagnets usually 
requires only a small set of characteristic parameters.
For example, a ferromagnet which possesses an uniaxial anisotropy can be 
modeled by a classical micromagnetic energy functional 
$E[{\bf \hat n}({\bf r})]$,
\begin{equation}
   E[{\bf \hat n}({\bf r})] = E_0 + \int d^3 r \,
        \left[ K \sin^2 \theta + A \left( {\bf \nabla} {\bf \hat n} \right)^2 
        \right] \, ,
\label{uniaxial}
\end{equation}
where ${\bf \hat n}({\bf r})$ is the unit vector that specifies the 
(space-dependent) local Mn spin orientation, and $\theta$ is its angle with 
respect to the easy axis. 
The dependence on the orientation of the magnetic moment is parametrized by the
anisotropy constant $K$.
The exchange constant (or spin stiffness) $A$ governs the energy cost to twist 
the orientations of adjacent spins relative to each other.
Together with the saturation moment $\mu_0 M_s$ and the magnetostatic energy, 
dropped for convenience in Eq.~(\ref{uniaxial}), the parameters $K$ and $A$ 
determine a whole variety of magnetic properties such as\cite{Skomski} 
domain-wall width $\delta_{\rm B}$, domain-wall energy per area $\gamma$, 
exchange length $l_{\rm ex}$, hardness parameter $\kappa$, single-domain 
radius $R_{\rm sd}$, and anisotropy field $\mu_0 H_0$.
In addition they determine the energy cost $\Omega_{k}$ of collective 
long-wavelength spin excitations, spin waves.
For uniaxial anisotropy the quantized energy is
\begin{equation}
   \Omega_{k} = {2K \over M_s/(g \mu_B)} + {2A \over M_s/(g \mu_B)} k^2 \, ,
\end{equation}
at wavevector $k$. 
Here, $g$ is the $g$-factor and $\mu_B$ the Bohr magneton.
There is an energy gap which is determined by the anisotropy $K$.
The spin exchange constant $A$ defines the curvature of the spin-wave 
dispersion.

The purpose of this paper is to present a microscopic theory for the
phenomenological parameters which characterize (Ga,Mn)As as a ferromagnet.
The form of the micromagnetic energy functional appropriate to the symmetry of 
the crystal will be addressed in Sec.~\ref{Section_classical}.
We expect that these predictions will be useful in interpreting experimental 
studies of these semiconductor's magnetic properties.
The search for the carrier density, Mn concentration and (III,Mn)V compound
semiconductor material for which the critical temperature has its maximum value
is, perhaps, the most important endeavor in this research area.
It has been guided to date by mean-field-theory%
\cite{Dietl97,Jungwirth99,Dietl00,Lee00,Abolfath01,Dietl00.2}
considerations that neglect the low-energy correlated magnetization 
fluctuations characterized by $M_s$, $K$, and $A$.
However, as we have emphasized earlier,%
\cite{Koenig1,Koenig2,Koenig3,Schliemann00.1} 
small exchange can also limit the temperature at which long-range magnetic 
order exists.

In recent work\cite{Koenig1,Koenig2,Koenig3} we developed a theory of 
diluted magnetic semiconductor ferromagnetism which accounts for dynamic 
correlations in the ordered state.
For simplicity we used parabolic bands for the itinerant carriers, leading to
isotropic ferromagnetism.
We found, in addition to the usual spin-wave mode, a continuum of Stoner 
excitations, and another collective branch of excitations, optical spin waves.
As we have shown,\cite{Koenig1,Schliemann00.1} the low-energy Goldstone modes 
can suppress the critical temperature in comparison to mean-field estimates 
and can even change trends in $T_c$ as a function of the system's parameters.

To address anisotropy one has to go beyond a parabolic-band model and take 
details of the band structure into account.
The itinerant-carrier bands are $p$-type and reflect the crystal symmetry of 
the underlying lattice.
Due to spin-orbit coupling the spin degrees of freedom also feel the crystal
anisotropy.
We will use a six-band description based on the Kohn-Luttinger Hamiltonian to
model the (Ga,Mn)As valence bands.
States near the Fermi energy in our model include substantial mixing with the 
split-off valence band making the six-band model we employ the minimal band 
model.
Mean-field calculations\cite{Dietl00,Abolfath01,Dietl00.2} as well as Monte 
Carlo studies\cite{Schliemann00.2} 
based on this more realistic band structure have been performed recently to 
address magnetic anisotropy effects and explore trends in the critical 
temperature.

In Sec.~\ref{Section_Hamiltonian} we set the starting point of our theory by 
deriving a formal expression for the effective action for the Mn impurity 
spins after integration out the itinerant carriers.
This formal development generalizes earlier work\cite{Koenig1,Koenig2,Koenig3}
to general spin-orbit coupled band models.
Using the effective action we numerically determine the zero-temperature 
spin-wave dispersion in Sec.~\ref{Section_SWT}.
Then, in Sec.~\ref{Section_classical} we establish the connection between our
spin-wave results and the classical micromagnetic energy functional, adjusted 
to the symmetry defined by the crystal structure.
Finally, in Sec.~\ref{Section_results}, we present numerical results for the
spin-wave dispersion and the exchange constant.
We find that for the parameter range of interest the exchange constant is 
enhanced by up to an order of magnitude compared to naive results obtained 
earlier, in which light-hole bands were neglected and the heavy-hole bands were 
approximated as parabolic.

\section{Hamiltonian and effective action}
\label{Section_Hamiltonian}

Our theory is based on the following model.
Magnetic ions with spin $S=5/2$ at positions ${\bf R}_I$ are 
antiferromagnetically coupled to valence-band carriers described by an 
envelope-function approach,
\begin{equation}
  H = H_0 + J_{\rm pd} \int \hspace{-1mm}d^3 r \;
  {\bf S}({\bf r}) \cdot {\bf s}({\bf r}) ,
\end{equation}
where ${\bf S}({\bf r}) = \sum_I {\bf S}_I \delta( {\bf r} - {\bf R}_I)$ is 
the impurity-spin density and $J_{\rm pd}>0$.
We approximate the impurity-spin density by a continuous functions (instead of 
a sum of delta functions).
Provided that $J_{\rm pd}$ is not strong enough to localize valence-band 
carriers near the Mn sites and the average distance between two Mn ions is 
small in comparison to the Fermi wavelength of the itinerant carriers, this 
approximation can be justified.
The itinerant-carrier spin density is expressed in terms of 
carrier field operators by ${\bf s}({\bf r})= \sum_{ij} \Psi_i^\dagger 
({\bf r}) {\bf s}_{ij} \Psi_j({\bf r})$ where $i$ and $j$ label the basis in 
Hilbert space of spin and orbital angular momentum, and ${\bf s}_{ij}$ is the 
corresponding representation of the spin operator.
The envelope-function Hamiltonian $H_0$ for the valence bands can be 
parametrized by a small number of symmetry-adapted parameters, the Luttinger
parameters $\gamma_1$, $\gamma_2$, $\gamma_3$, and the spin-orbit coupling
$\Delta_{\rm so}$ which splits the six states at the band edge into a quartet 
and a doublet. 
For explicit expressions of $H_0$ as well as representations of the spin 
matrices in a coordinate system in which the $x$, $y$, and $z$-axis 
are along the crystal axes, we refer to Eqs.~(A8)-(A10) of 
Ref.~\onlinecite{Abolfath01}, based on the Kohn-Luttinger Hamiltonian.%
\cite{Luttinger55}

Phenomenological Hamiltonians of this form have proven successful
in understanding optical properties of the closely related magnetically
doped II-VI semiconductors.\cite{Furdyna,Dietl94}  
In that case however, the inclusion of direct antiferromagnetic interactions 
between Mn spins that lie on neighboring lattice sites proved to be important.  
This interactions appear to be much weaker in the III-V case, although
this difference is not fully understood.\cite{Dietl00.2}
We do not include these spin-spin interactions in our calculations, because we 
have no knowledge of the strength of the coupling coefficient.  
If included, they would mainly influence short-distance spin excitations which 
are not in any event the focus of this work.

We note that local-spin-density-approximation electronic structure
calculations,\cite{Sanvito00} taken at face value, find a strong hybridization 
between $p$ and $d$ carriers, which is not completely consistent with the 
model we use.
However, given the overwhelming success\cite{Furdyna,Dietl94} of the present 
model in the case of paramagnetic (II,Mn)VI materials, we find it unlikely that
the Mn $d$ electrons are itinerant.
If they were, the phenomenological approach we study here would be incomplete.

Our treatment of the Mn spin system as a continuum greatly simplifies our
calculation by eliminating disorder associated with randomness in the Mn 
positions.
We do anticipate that the disorder can influence both anisotropy and exchange
constant at the low- and high-carrier-concentration extremes.
These effects are, however, outside the scope of the present study.
We will identify several important features of the microscopic physics in the
disorder-free model that we expect to be robust.

Our first goal is to integrate out the itinerant carriers and arrive at an
effective description for the impurity-spin degrees of freedom.
This procedure follows the analysis presented in Ref.~\onlinecite{Koenig1} for
the (simpler) two-band model.
For small fluctuations around its mean-field polarization, we can approximate
the spin operators by $S^+ ({\bf r}) \approx b({\bf r}) \sqrt{2N_{\rm Mn}S}$,
$ S^- ({\bf r})  = \left( S^+ ({\bf r}) \right)^\dagger$, and 
$S^z ({\bf r}) = N_{\rm Mn}S - b^{\dag}({\bf r}) b({\bf r})$, where  
$b^{\dag}({\bf r}), b({\bf r})$ are bosonic Holstein-Primakoff\cite{Auerbach94} 
fields.
The quantization axis $z$ is chosen here along the zero-temperature spin 
orientation.
After integrating out the itinerant carriers the partition function,
$Z = \int {\cal D} [\bar z z] \exp (- S_{\rm eff} [\bar z z])$, is governed by
the effective action for the impurity spins, 
\begin{equation}
  S_{\rm eff} [\bar z z] =  S_{\rm BP} [\bar z z]
  - \ln \det \left[ (G^{\rm MF})^{-1} + \delta G^{-1}(\bar zz) \right] \, ,
\label{effective action}
\end{equation}
where $S_{\rm BP} [\bar z z] = \int_0^\beta d\tau \int d^3 r \, \bar z 
\partial_\tau z$ is the usual Berry's phase term, and the complex numbers
$\bar z$ and $z$ label the bosonic degrees of freedom.
In Eq.~(\ref{effective action}), we have already split the total kernel 
$G^{-1}$ into a mean-field part $(G^{\rm MF})^{-1}$ and a fluctuating part 
$\delta G^{-1}$,
\begin{eqnarray}
   (G^{\rm MF})^{-1}_{ij} &=&
 \left( \partial_\tau -\mu \right)\delta_{ij} + \langle i| H_0 | j \rangle
        + N_{\rm Mn}J_{\rm pd}S s^z_{ij}
\\
   \delta G^{-1}_{ij}(\bar zz) &=& {J_{\rm pd}\over 2} \left[ \left( 
        z s^-_{ij} + \bar z s^+_{ij} \right) \sqrt{2N_{\rm Mn}S} 
        - 2 \bar z z s^z_{ij} \right]
\end{eqnarray}
where $\mu$ denotes the chemical potential, and $i$ and $j$ range over a 
complete set of hole-band states.
In the following we define the mean-field energy 
$\Delta = N_{\rm Mn}J_{\rm pd}S$ to flip the spin of an itinerant carrier.
The physics of the itinerant carriers is embedded in the effective action
of the magnetic ions.
It is responsible for the retarded and non-local character of the interactions 
between magnetic ions.

\section{Independent spin-wave theory}
\label{Section_SWT}

Independent spin-wave theory is obtained by expanding 
Eq.~(\ref{effective action}) up to quadratic order in $z$ and performing
Matsubara imaginary time and space Fourier transforms.
Since $\delta G^{-1}$ is at least linear in $z$ the series 
\begin{equation}
  S_{\rm eff} [\bar z z] = \sum_{n=0}^\infty S^{(n)}_{\rm eff}[\bar zz] \, ,
\label{series}
\end{equation}
can be truncated after $n=2$, where $n$ denotes the order in $\delta G^{-1}$.
The zeroth-order contribution,
$ S^{(0)}_{\rm eff} [\bar z z] = 
S_{\rm BP}[\bar zz] - \ln \det (G^{\rm MF})^{-1}$
contains the Berry's phase term
\begin{equation}
  S_{\rm BP}[\bar zz] = \sum_{m,{\bf k}} (-i \nu_m) 
        \bar z({\bf k},\nu_m) z({\bf k},\nu_m) \, ,
\label{SB}
\end{equation}
and the mean-field contribution from the itinerant carriers, which is 
independent of the bosonic fields $z$ and $\bar z$.
Here, $\nu_m$ are the bosonic Matsubara frequencies.

The next term of the expansion, 
$S^{(1)}[\bar zz] =  - {\rm tr} \left( G^{\rm MF} \delta G^{-1} \right)$, 
reads in Fourier representation
\begin{eqnarray}
   S^{(1)}_{\rm eff}[\bar zz] &=& 
        {J_{\rm pd} \over (\beta V)^2} \sum_{n,{\bf q}} \sum_{ij} \, 
        G^{\rm MF}_{ij} ({\bf q},\omega_n) s^z_{ji}
\nonumber \\ && \times
        \sum_{m,{\bf k}} \bar z({\bf k},\nu_m) z({\bf k},\nu_m) \, ,
\end{eqnarray}
plus terms linear in $z$ and $\bar z$.
Here, $\omega_n$ and $\nu_m$ are fermionic and bosonic Matsubara frequencies,
respectively.
To determine the mean-field Green's functions we diagonalize the matrix
$\langle i| H_0({\bf q}) + \Delta s^z | j \rangle$ for each wavevector 
${\bf q}$ and denote eigenvalues and eigenstates by
$\epsilon_{\alpha}({\bf q})$ and $|\alpha \rangle$, respectively.
Since the itinerant-carrier spin density
$\langle {\bf s} \rangle = (1/V) \sum_{\bf q} \sum_{\alpha} 
\, f [\epsilon_\alpha({\bf q})] \langle \alpha | {\bf s} | \alpha \rangle$
is aligned antiparallel to the impurity spin-polarization axis (otherwise this 
would not be an easy axis), the terms linear in $z$ and
$\bar z$ drop out, and we get
\begin{equation}
   S^{(1)}_{\rm eff}[\bar zz] =
        {J_{\rm pd} p\xi \over 2\beta V}
        \sum_{m,{\bf k}} \bar z({\bf k},\nu_m) z({\bf k},\nu_m) \, ,
\label{S1}
\end{equation}
where $\xi = -2 \langle s^z \rangle /p$ is the fractional 
itinerant-carrier polarization, $0\le \xi \le 1$.

For the second-order term of the expansion, 
$S^{(2)}_{\rm eff}[\bar zz] = {1\over 2}\, {\rm tr} \left( G^{\rm MF} 
\delta G^{-1} G^{\rm MF} \delta G^{-1} \right)$,
we find 
\begin{eqnarray}
   S^{(2)}_{\rm eff}[\bar zz] &=& {N_{\rm Mn} J_{\rm pd}^2 S \over 4\beta V^2}
        \sum_{m,{\bf q},{\bf k}} \sum_{\alpha\beta}
        { f [\epsilon_\alpha({\bf q})] - 
          f [\epsilon_\beta({\bf q}+{\bf k})] \over 
          i\nu_m + \epsilon_\alpha({\bf q})  
                - \epsilon_\beta({\bf q}+{\bf k}) }
\nonumber \\ &&
        \left[
        2 s^+_{\alpha\beta} s^-_{\beta\alpha} 
         \bar z({\bf k},\nu_m) z({\bf k},\nu_m)
\nonumber \right. \\ && \left.
        + s^+_{\alpha\beta} s^+_{\beta\alpha}  
        \bar z({\bf k},\nu_m) \bar z(- {\bf k},-\nu_m)
\nonumber \right. \\ && \left.
        + s^-_{\alpha\beta} s^-_{\beta\alpha}  
        z({\bf k},\nu_m) z(- {\bf k},-\nu_m)
        \right] + {\cal O}\left( z^3 \right) \, ,
\label{S2}
\end{eqnarray}
with $s^\pm_{\alpha\beta} = \langle \alpha | s^\pm | \beta \rangle$ and  
$s^\pm_{\beta\alpha} = \langle \beta | s^\pm | \alpha \rangle$.
Note, that the indices $\alpha$ and $\beta$ label the single-particle 
eigenstates for valence-band carriers with {\it different} wavevectors, namely
{\bf q} and {\bf q}+{\bf k}, respectively.

The Matsubara frequency $\nu_m$ in the denominator on the r.h.s of 
Eq.~(\ref{S2}) accounts for the dynamics of the itinerant carriers.
This frequency dependence is crucial to account for the existence of the 
Stoner spin-flip continuum and the optical spin-wave mode.
On the other hand, the existence of the usual spin wave follows already from
the static limit (i.e., when the frequency dependence in the denominator
in Eq.~(\ref{S2}) is dropped), and the spin-wave dispersion is described rather
accurately.

The sum of Eqs.~(\ref{SB}), (\ref{S1}), and (\ref{S2}) is a quadratic form in 
the bosonic fields $z$ and $\bar z$.
The zeros of the kernel define the spin-wave energies $\Omega_{\bf k}$ as a
function of momentum ${\bf k}$ (after analytic continuation 
$i\nu_m \rightarrow \Omega +i0^+$).
In the following we go the static limit as discussed above.
We define the quantities
\begin{equation}
        E_{\bf k}^{\sigma \sigma'}  = - {1\over V} \sum_{\bf q} 
        \sum_{\alpha\beta}
        { f [\epsilon_\alpha({\bf q})] - 
          f [\epsilon_\beta({\bf q}+{\bf k})] \over
          \epsilon_\alpha({\bf q}) - \epsilon_\beta ({\bf q}+{\bf k}) } 
        s^\sigma_{\alpha\beta} s^{\sigma'}_{\beta\alpha}
\label{E}
\end{equation}
with $\sigma,\sigma'=\pm$, and perform a Bogoliubov transformation, which
eventually yields
\begin{equation}
  {\Omega_{\bf k}\over \Delta} = { J_{\rm pd} \over 2}
        \sqrt{ \left( { p\xi \over \Delta }
        - E_{\bf k}^{+-} \right)^2
        - \left|E_{\bf k}^{++}\right|^2 } \, .
\label{dispersion}
\end{equation}
From the definition Eq.~(\ref{E}) we see that $E_{\bf k}^{+-}$ is real and 
$E_{\bf k}^{--}= \left( E_{\bf k}^{++}\right)^*$.
Equation~(\ref{dispersion}) is the central result of this section.
The remaining task is to evaluate the fractional itinerant-carrier 
polarization $\xi$ and the quantities $E_{\bf k}^{+-}$ and $E_{\bf k}^{++}$
numerically.

Before we carry on with establishing the relation between 
Eq.~(\ref{dispersion}) and micromagnetic parameters of a classical energy 
functional, we make three remarks:

(i) Correlation effects among the Mn spins, which are not described by the 
mean-field picture, enter our theory via the contribution 
$S^{(2)}_{\rm eff}[\bar zz]$. 
To reduce our theory to the mean-field level we would have to neglect this 
term, i.e., truncate the series Eq.~(\ref{series}) already after $n=1$.
In this case, the energy $\Omega^{\rm MF} = J_{\rm pd} p\xi/2$ of a Mn spin 
excitation would be dispersionless and by a factor of $p\xi/(2N_{\rm Mn}S)$
smaller than the mean-field energy $\Delta$ to flip an itinerant-carrier spin.
Due to correlations between Mn and band-spin orientations, however, the 
spin-wave energy $\Omega_{\bf k}$ is always smaller than $\Omega^{\rm MF}$.

(ii) In the absence of spin-orbit coupling all products of the form 
$s^+_{\alpha\beta} s^+_{\beta\alpha}$ or  
$s^-_{\alpha\beta} s^-_{\beta\alpha}$ vanish.
As a consequence, $E_{\bf k}^{++}=E_{\bf k}^{--}=0$ for all ${\bf k}$.
This statement is even true for finite spin-orbit coupling in case when the 
valence bands are isotropic.

(iii) To go beyond the static limit, we may expand the fraction on the r.h.s of
Eq.~(\ref{S2}) up to linear order in $i\nu_m$.
As shown in Appendix~\ref{append_berry}, this linear correction simply amounts 
to a renormalization of the Berry's phase term by replacing 
$\Omega \rightarrow \Omega (1-x)$.
In the absence of spin-orbit coupling we find that $x$ is just the ratio of
the spin densities, $x = \langle s^z \rangle /(N_{\rm Mn}S)$, where
$\langle \ldots \rangle = (1/V) \sum_{\bf q} \sum_\alpha 
f[\epsilon_\alpha({\bf q})] \, \langle \alpha| \ldots | \alpha \rangle$.
For finite spin-orbit coupling but with a band Hamiltonian that is invariant
under rotation in space, the spin $s^z$ has to be replaced by the total 
angular momentum $s^z + l^z$, i.e., we find 
$x = \langle s^z+l^z \rangle /(N_{\rm Mn}S)$.

The renormalization factor $(1-x)$ indicates that in a semiclassical
picture, as employed in Sec.~\ref{Section_classical}, the effective spin 
density is not quite given by the Mn impurities, but has to be reduced due to 
coupling to the valence-band carriers.  
On the other hand, $x$ is always small since the impurity spin density 
$N_{\rm Mn}$ is larger than the itinerant-carrier concentration $p$ and the 
Mn spin $S=5/2$ is comparatively large.
We, therefore, stick to the static limit in the following discussion.

\section{Easy axis, energy gap and spin stiffness}
\label{Section_classical}

The purpose of this section is to establish the connection between the 
spin-wave dispersions evaluated later and the micromagnetic energy functional.
The non-local magnetostatic contribution which is omitted from our theory can 
be added as needed in applications.
The short-range part of the functional is a symmetry-adapted gradient expansion
of the energy density $e[{\bf \hat n}]$.
In magnetism literature the non-constant portion of the zeroth-order term 
$e^{\rm ani}[{\bf \hat n}]$ is known as the magnetic anisotropy energy, and the
leading gradient term $e^{\rm ex}[{\bf \hat n}]$ is known as the exchange 
energy.
As shown in Ref.~\onlinecite{Abolfath01} magnetic anisotropy effect in 
ferromagnetic semiconductors are, in the absence of strain, very well 
described by a cubic harmonic expansion which is truncated after sixth order,
\begin{equation}
\label{aniso}
   e^{\rm ani}[{\bf \hat n}] = 
        K_1 \left( n_x^2 n_y^2 + n_x^2 n_z^2 + n_y^2 n_z^2 \right)
        + K_2 \left( n_x n_y n_z \right) ^2 
\end{equation}
with anisotropy parameters $K_1$ and $K_2$.
Correlation of spin polarizations at different positions are described by the
gradient term $e^{\rm ex}[{\bf \hat n}]$.
In order to address long-wavelength spatial fluctuations, we expand the 
gradient term up to lowest nonvanishing order,
\begin{equation}
\label{gradient}
   e^{\rm ex}[{\bf \hat n}] = 
        \sum_{a,b \in \{x,y,z\}} A_{ab} |\partial_a n_b|^2 \, ,
\end{equation}
with exchange parameters $A_{ab}$.
We find in our numerical calculations that anisotropy in the exchange 
constant is negligibly small, i.e., we can choose $A_{ab} = A$ for all $a,b$,
as generally assumed in the magnetism literature.

\begin{figure}
\centerline{\includegraphics[width=8.cm]{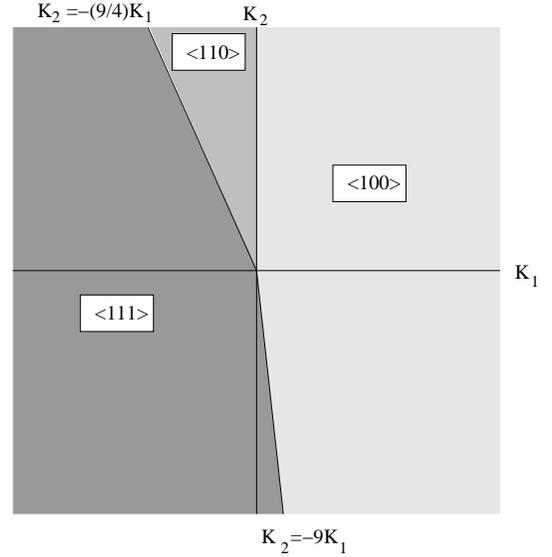}}
\caption{Easy-axis direction as a function of $K_1$ and $K_2$.}
\label{fig1}
\end{figure}

To establish the connection of the energy functional to our microscopic 
spin-wave calculation, we first have to determine the direction of the 
mean-field spin polarization and then to study small fluctuations.
The first step is achieved by minimizing the energy Eq.~(\ref{aniso}).
It is easy to show that the mean-field orientation ${\bf \hat n}^{\rm MF}$
can only point along a high-symmetry axis $\langle 100 \rangle$,
$\langle 110 \rangle$, $\langle 111 \rangle$, or an equivalent direction
(except for the special case $K_2=0$ and $K_1>0$, where we find an easy-plane 
anisotropy in the planes $n_x=0$, $n_y=0$ or $n_z=0$, and, of course, the 
isotropic case $K_1=K_2=0$).
In Fig.~\ref{fig1} we show how the mean-field polarization direction depends 
on $K_1$ and $K_2$.

Now we consider small fluctuations around the mean-field orientation 
${\bf \hat n}^{\rm MF}$.
The kernel of the quadratic form 
$e^{\rm ani}[{\bf \hat n}] - e^{\rm ani}[{\bf \hat n}^{\rm MF}]$ has two 
eigenvalues $\lambda_1$ and $\lambda_2$.
We find for ${\bf \hat n}^{\rm MF}$ along $\langle 100 \rangle$ that 
$\lambda_1 = \lambda_2 = K_1$, for $\langle 110 \rangle$ we get
$\lambda_1 = -K_1$ and $\lambda_2 = (2K_1+K_2)/4$, and for 
$\langle 111 \rangle$ we obtain $\lambda_1 = \lambda_2 = - (6K_1+2K_2)/9$.
In all cases, $\lambda_1$ and $\lambda_2$ are positive.
Quantizing the collective spin coordinate at long wavelengths, it follows that
\begin{equation}
   \Omega_{k} = {2\sqrt{\lambda_1 \lambda_2}\over N_{\rm Mn}S} 
        +  {2A\over N_{\rm Mn}S} \, k^2 + {\cal O}(k^4) \, .
\label{gap+stiffness}
\end{equation}
There are two alternative ways to determine the energy gap $\Omega_{k=0}$.
One can either perform the $k=0$ limit of the spin-wave dispersion 
Eq.~(\ref{dispersion}) or evaluate the coefficients $\lambda_1$ and $\lambda_2$
from calculating the energy for mean-field orientation 
${\bf \hat n}^{\rm MF}$ along the three high-symmetry axes 
$\langle 100 \rangle$, $\langle 110 \rangle$, and $\langle 111 \rangle$. 
The numerical effort of the latter procedure, which has been used previously in
Ref.~\onlinecite{Abolfath01}, is lesser for given accuracy.
The virtue of the spin-wave calculation is to determine the exchange constant 
$A$.

We conclude this section with two remarks:

(i) In case that the easy axis is along $\langle 100 \rangle$ or 
$\langle 111 \rangle$ (as it is for all parameter sets considered in 
Sec.~\ref{Section_results}), the energy cost of tilting the polarization axis 
by small angles is independent of the direction of the deflection. 
This is required by symmetry and indicated by the fact that $\lambda_1$ equals
$\lambda_2$.
As shown in Appendix~\ref{append_gap}, the term $E^{++}_{k=0}$ then vanishes,
and the $k=0$ limit of Eq.~(\ref{dispersion}) is identical to the energy
gap calculated from standard perturbation theory where the perturbation 
describes the deviation of the spin polarization from the mean-field direction.

(ii) The denominators $N_{\rm Mn}S$ in Eq.~(\ref{gap+stiffness}) corresponds 
to the static limit employed in our calculation. 
The renormalization of the Berry's phase term due to corrections to the static
limit (see discussion in the previous section) could be accounted for by 
multiplying $N_{\rm Mn}S$ with $(1-x)$.

\section{Numerical results for the spin-wave dispersion}
\label{Section_results}

In this section we present numerical results for the spin-wave dispersion of
(Ga,Mn)As.
From these calculations we extract the spin stiffness as a function of the 
itinerant-carrier concentration $p$ and on the exchange coupling $J_{\rm pd}$.
To model the sample which showed the highest transition temperature of 
$110 {\rm K}$ so far, we choose as parameters\cite{Omiya00} 
$N_{\rm Mn} = 1 \, {\rm nm}^{-3}$, $p = 0.35 \, {\rm nm}^{-3}$, and 
$J_{\rm pd} = 0.068 \, {\rm eV \, nm}^{-3}$.
As a consequence, the mean-field spin-splitting gap for the itinerant carriers
is $\Delta =  N_{\rm Mn} J_{\rm pd} S = 0.17  \, {\rm eV}$.

\subsection{Isotropic vs. six-band model}

The origin of ferromagnetism, the nature of the spin excitations, and trends
in the critical temperatures can be explained within a simple model which 
describes the itinerant carriers by parabolic bands.%
\cite{Koenig1,Koenig2,Koenig3,Lyu01}
For more quantitative statements, a more realistic description of the band 
structure should be employed.

\begin{figure}
\centerline{\includegraphics[width=8.cm]{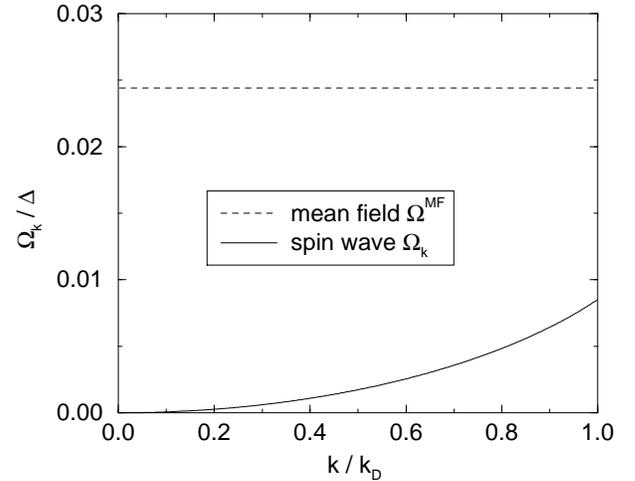}}
\caption{Spin-wave dispersion for the isotropic model for itinerant-carrier
        density $p=0.35 \, {\rm nm}^{-3}$, impurity-spin concentration
        $N_{\rm Mn} = 1.0 \, {\rm nm}^{-3}$ and exchange coupling 
        $J_{\rm pd} = 0.068 \, {\rm eV \, nm}^{-3}$ (which yields 
        $\Delta = 0.17 \, {\rm eV}$).}
\label{fig2}
\end{figure}

In Fig.~\ref{fig2} we show results for the isotropic model with two parabolic 
band with effective mass $m^* = 0.5 m_e$ (a Debye cutoff $k_D$ with
$k_D^3 = 6\pi^2 N_{\rm Mn}$ ensures that we include the correct number of 
magnetic ion degrees of freedom).
We find for the majority-spin Fermi energy $\epsilon_F = 0.44 \, {\rm eV}$ 
measured from the bottom of the band.
This yields the itinerant-carrier polarization $\xi = 0.35$.
The dashed line in Fig.~\ref{fig2} marks the mean-field spin-flip energy,
obtained by neglecting correlation. 
Since the spin-wave energies are far below the mean-field result, the 
isotropic model suggest that for these parameters correlation is very 
important and the critical temperature is limited by collective fluctuations.
According to the classification scheme introduced in 
Ref.~\onlinecite{Schliemann00.1}, the system would be in the 
''RKKY collective regime''.

By fitting a parabola at small momenta we find 
$\Omega_k/\Delta = 0.0068 (k/k_D)^2$ which yields
$A = 0.095 \, {\rm meV \, nm}^{-1} = 0.015 \, {\rm pJ \, m}^{-1}$.

Now we use the six-band Kohn-Luttinger Hamiltonian with Luttinger parameters
$\gamma_1 = 6.85$, $\gamma_2 = 2.1$, and $\gamma_3 = 2.9$ and spin-orbit 
coupling $\Delta_{\rm so} = 0.34 \, {\rm eV}$.
In the absence of the Mn ion, this model has anisotropic heavy- and light-hole
bands with masses $m_h\approx 0.498m_e$ and $m_l \approx 0.086m_e$. 
The spirit of the naive parabolic band model is the hope that only the band
with the larger density of states matters and the anisotropy is unimportant.
We will see that these hopes are not fulfilled.

We find a much lower Fermi energy $\epsilon_F = 0.25 \, {\rm eV}$ than 
obtained for the isotropic model, and, therefore, a much higher 
itinerant-carrier spin polarization $\xi = 0.73$.
By calculating the mean-field energy for a Mn spin polarization along the
three high-symmetry axes we determine the anisotropy parameters as
$K_1 = 19.6 \times 10^{-6} \, {\rm eV \, nm}^{-3}$ and
$K_2 = 1.6 \times 10^{-6} \, {\rm eV \, nm}^{-3}$.
As a consequence, the easy axis is $\langle 100 \rangle$ and the energy gap,
$\Omega_{k=0}/\Delta = 9.2 \times 10^{-5}$, is very small in comparison to the 
bandwidth of the spin-wave dispersion.

\begin{figure}
\centerline{\includegraphics[width=8.cm]{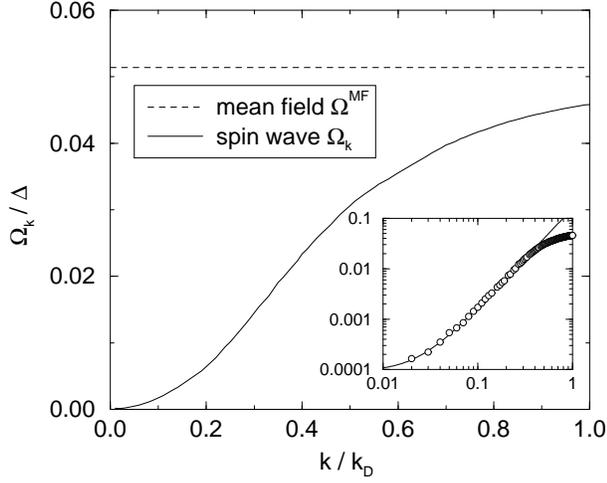}}
\caption{Main panel: Spin-wave dispersion for the 6-band model for 
        itinerant-carrier density $p=0.35 \, {\rm nm}^{-3}$, impurity-spin 
        concentration $N_{\rm Mn} = 1.0 \, {\rm nm}^{-3}$ and exchange 
        coupling $J_{\rm pd} = 0.068 \, {\rm eV \, nm}^{-3}$ (which yields 
        $\Delta = 0.17 \, {\rm eV}$).
        The momentum $k$ is chosen to be parallel to the easy axis 
        $\langle 100 \rangle$.
        Inset: Spin-wave dispersion on a log-log plot (circles) and the 
        fit $9.2\times 10^{-5} + 0.16 (k/k_D)^2$ (solid line).}
\label{fig3}
\end{figure}

In Fig.~\ref{fig3} we show the spin-wave dispersion for wavevectors ${\bf k}$
along the easy axis.
We observe that the effect of $E_{\bf k}^{++}$ in Eq.~(\ref{dispersion}) is
negligibly small and can, therefore, be dropped.
Furthermore, we find that the dispersion for a spin wave perpendicular to the 
easy axis cannot be distinguished from the dispersion of spin waves along the
easy axis within numerical accuracy.
By fitting a parabola at small momenta we find 
$\Omega_{k}/\Delta = \Omega_{k=0}/\Delta + 0.16 (k/k_D)^2$ which 
yields $A = 2.2 \, {\rm meV \, nm}^{-1} = 0.36\,{\rm pJ \, m}^{-1}$
(see inset of Fig.~\ref{fig3}).
Furthermore, we see that the energy gap $\Omega_{k=0}/\Delta$ determined in the
previous paragraph is consistent with our spin-wave results. 
Employing the six-band model we see that for the given parameters the spin-wave
energies are much closer to the mean-field value than for the isotropic model.
According to our classification given in Ref.~\onlinecite{Schliemann00.1}
the system is rather in the ''mean-field regime'', which explains the success 
of mean-field theory to reproduce the critical temperature.

\subsection{Six-band model plus strain}

The lattice constants of (Ga,Mn)As and GaAs do not match.
Since low-temperature molecular beam epitaxy (MBE) has to be used to overcome 
the low solubility of Mn in GaAs, even thick films of (Ga,Mn)As grown on GaAs 
cannot relax to their equilibrium.
The lattice of (Ga,Mn)As is instead locked to that of the underlying substrate.
This induces strain which breaks the cubic symmetry.
The influence of MBE growth lattice-matching strain on hole bands of cubic 
semiconductors is well understood.\cite{Chow99,Bir74}
This effect can easily be accounted for adding a strain term to the Hamiltonian
(we use Eq.~(32) of Ref.~\onlinecite{Abolfath01} with strain parameters
$e_0 = -0.0028$ and $\Gamma = -3.24 \, {\rm eV}$).

We choose the growth direction to be along $\langle 001 \rangle$ and compute 
the energy for five different directions $\langle 100 \rangle$, 
$\langle 001 \rangle$, $\langle 110 \rangle$, $\langle 011 \rangle$, and 
$\langle 111 \rangle$.
The lowest energy is found for $\langle 100 \rangle$, i.e., we find an 
easy-axis anisotropy where the easy axis is in the plane perpendicular to the 
growth direction, in accordance with experiments.\cite{Ohno.proc}
From the expansion of the mean-field anisotropy energy for small fluctuations
around $\langle 100 \rangle$ and determining the eigenvalues for small 
fluctuations we estimate that the energy gap in the dispersion is larger than in 
the absence of strain by a factor of 1.4, i.e., still small.
The spin stiffness derived from the curvature of the spin-wave dispersion
is identical to that in the absence of strain within the accuracy of our 
numerical calculations.
We will, therefore, ignore the effect of strain in the following.

\subsection{Spin stiffness}

\begin{figure}
\centerline{\includegraphics[width=8.cm]{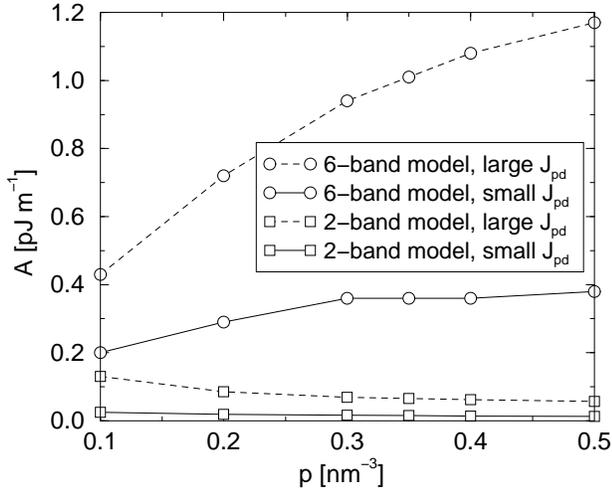}}
\caption{Exchange constant $A$ as a function of itinerant-carrier density $p$
        for the six-band and the two-band model for two different values of
        $J_{\rm pd} = 0.068 \, {\rm eV \, nm}^{-3}$ (solid lines) and 
        $0.136 \, {\rm eV \, nm}^{-3}$ (dashed lines).
        The impurity-spin concentration is chosen as 
        $N_{\rm Mn} = 1.0 \, {\rm nm}^{-3}$, which yields
        $\Delta = 0.17 \, {\rm eV}$ (solid lines) and 
        $\Delta = 0.34 \, {\rm eV}$ (dashed lines), respectively.}
\label{fig4}
\end{figure}

In Fig.~\ref{fig4} we show the exchange constant (or spin stiffness) $A$ as a 
function of the itinerant-carrier density for both the isotropic two-band and 
the full six-band model.
We find that the spin stiffness is much larger for the six-band calculation 
than for the two-band model.
Furthermore, for the chosen range of itinerant-carrier densities the trend is 
different: in the two-band model the exchange constant decreases with 
increasing density, while for the six-band description we observe an increase
with a subsequent saturation.

Experimental estimates for $J_{\rm pd}$ vary from $0.054 \, {\rm eV \, nm}^3$ 
to $0.15 \, {\rm eV \, nm}^3$, with more recent work suggesting a value toward 
the lower end of this range.\cite{Ohno98.2,Okabayashi98,Omiya00}
To address the dependence of the spin stiffness on $J_{\rm pd}$, we show in
Fig.~\ref{fig4} results for a two values of $J_{\rm pd}$ which differ by a factor
two.

To understand the different behavior of the two-band and the six-band model
we recall\cite{Koenig2,Koenig3} that the two-band model predicts 
$A = p\hbar^2 /(8m^*)$ for low densities $p$, while at high densities $A$ 
decreases as a function of $p$ (note that in Refs.~\onlinecite{Koenig2} and 
\onlinecite{Koenig3} the spin stiffness was characterized by $\rho = 2A$).
The crossover occurs near $\Delta \sim \epsilon_F$.
The difference in the trends seen for the two- and six-band model in 
Fig.~\ref{fig4} is consistent with the observation that, at 
given itinerant-carrier concentration $p$, the Fermi energy $\epsilon_F$ is 
much smaller when the six-band model is employed, where more bands are 
available for the carriers, than in the two-band case.
Furthermore, we emphasize that, even in the limit of low carrier concentration,
it is not only the (heavy-hole) mass of the lowest band which is important for 
the spin stiffness.
Instead, a collective state in which the spins of the itinerant carriers 
follow the spatial variation of a Mn spin-wave configuration will involve the 
light-hole band, too.

\section*{Conclusion}

In conclusion we present a microscopic calculation of micromagnetic 
parameters for ferromagnetic (Ga,Mn)As.
We draw the connection of the anisotropy and exchange constant of a
classical energy functional to the gap and curvature of the spin-wave 
dispersion.
Numerical results for the spin stiffness as a function of itinerant-carrier 
concentration and $p-d$ exchange coupling are shown.
We find that the energy gap is much smaller than the bandwidth, the spin 
stiffness is nearly isotropic, and strain does not effect the dispersion much.
Furthermore, we see that a model with isotropic valence bands underestimates 
the spin stiffness considerably.

\section*{Acknowledgements}

We acknowledge useful discussions with M.~Abolfath, B.~Beschoten, T.~Dietl,
J.~Furdyna, H.H.~Lin, and J.~Schliemann.
This work was supported by the Deutsche Forschungsgemeinschaft, the Ministry of 
Education of the Czech Republic, the EU COST Program, the Welch 
Foundation, the Indiana $21^{\rm st}$ Century Fund, and the Office of Naval 
Research and the Research Foundation of the State University of New York under 
grant number N000140010951.

\begin{appendix}

\section{Renormalization of Berry's phase term}
\label{append_berry}

In this section, we show that the linear correction beyond the static limit 
yields a correction to the Berry's phase.
This accounts for the fact that the effective spin in a semiclassical approach 
does not have the length $S$ of the Mn impurities by is reduced by a factor 
$(1-x)$ due to coupling to the itinerant-carrier spin degree of freedom.
We expand the r.h.s of Eq.~(\ref{S2}) up to linear order in $i\nu_m$.
The terms with $\bar z \bar z$ and $z z$ vanish since it is an odd
function under the following operation: we shift 
${\bf q} \rightarrow -{\bf q} - {\bf k}$, use the fact that 
$\epsilon_\alpha( -{\bf q} - {\bf k}) = \epsilon_\alpha( {\bf q} + {\bf k})$
and $\epsilon_\beta( -{\bf q}) = \epsilon_\beta( {\bf q})$ and, eventually, 
exchange $\alpha \leftrightarrow \beta$.
Hence, we only have to deal with the term which involves $\bar z z$.

Assume that there is an operator $Q$ such that $N_{\rm Mn}J_{\rm pd}S s^+ =
[H,Q]$ and $N_{\rm Mn}J_{\rm pd}S s^- = - [H,Q^\dagger]$.
Then the renormalization for the Berry's phase is given by $(1-x)$ with
\begin{equation}
   x = {1\over 2N_{\rm Mn}SV} \sum_{\bf q} \sum_{\alpha}
        f [\epsilon_\alpha({\bf q})]
        \langle \alpha | [Q^\dagger,Q] | \alpha \rangle 
\end{equation}
independent of ${\bf k}$.
In the absence of spin-orbit coupling we choose $Q=s^+$, which yields
\begin{equation}
   x = {\langle s^z \rangle \over N_{\rm Mn}S } = { p\xi \over 2N_{\rm Mn}S } 
         \, ,
\label{Berry_1}
\end{equation}
i.e., $x$ is just the ratio of itinerant-carrier spin concentration to Mn spin 
density.
For finite spin-orbit coupling the spin of the itinerant carrier is coupled to 
the orbital angular momentum.
If the band Hamiltonian is invariant under rotation in space, we can choose 
$Q=s^++l^+$ and find
\begin{equation}
   x = {\langle s^z+l^z \rangle \over N_{\rm Mn}S } \, .
\label{Berry_2}
\end{equation}
In this case the correction is given by the ratio of the total angular 
momentum density of the itinerant carriers and the Mn concentration.

If the valence bands are described by the Kohn-Luttinger Hamiltonian, however,
the total angular momentum is no longer a conserved quantity.
Or, equivalently, the orbital angular momentum of the valence-band carriers
couples to the crystal lattice, and the simple form Eq.~(\ref{Berry_2}) no
longer holds.

\section{Spin-wave dispersion gap}
\label{append_gap}

The goal of this section is to rederive the $k=0$ limit of the spin-wave 
energy Eq.~(\ref{dispersion}) by standard perturbation theory where the 
perturbation describes the deviation of the spin polarization from the 
mean-field direction $\langle 100 \rangle$ or $\langle 111 \rangle$.

\subsection{Proof that $E_{k = 0}^{++} = 0$ for $\langle 100 \rangle$ or 
$\langle 111 \rangle$ easy axis}

We start by showing that
\begin{equation}
        E_{k = 0}^{++}  = - {1\over V} \sum_{\bf q} 
        \sum_{\alpha\beta}
        { f [\epsilon_\alpha({\bf q})] - f [\epsilon_\beta({\bf q})] \over
          \epsilon_\alpha({\bf q}) - \epsilon_\beta ({\bf q}) } 
        s^+_{\alpha\beta} s^+_{\beta\alpha}
\label{E++0}
\end{equation}
vanishes, if the mean-field polarization is along $\langle 100 \rangle$ or 
$\langle 111 \rangle$.
Note that $\alpha$ and $\beta$ now label the {\it same} basis states.

Let ${\bf \tilde q}$ be a wavevector which is obtained from ${\bf q}$ by
rotation about the $z$-axis (which is defined by the mean-field Mn spin
polarization direction) by an angle $\varphi$ which respects the symmetry of 
the crystal.
If the $z$-axis is $\langle 100 \rangle$ or $\langle 111 \rangle$, then the 
allowed angles are $\varphi \in \{ 0,\pi/2,\pi,3\pi/2 \} $ or 
$\{ 0,2\pi/3,\pi,4\pi/3 \} $, respectively.
For $\langle 110 \rangle$ it would be  $\varphi \in \{ 0,\pi \} $.
Due to the symmetry of the crystal, the spectrum at ${\bf \tilde q}$, labeled
by $\tilde \alpha$ (or $\tilde \beta$), is identical to that at ${\bf q}$, 
i.e., $\epsilon_{\tilde \alpha} = \epsilon_{\alpha}$.
The corresponding eigenstates are connected by 
$|\tilde \alpha \rangle = U |\alpha \rangle$ with 
$U=\exp[i (s^z+l^z)\varphi]$.

Since the spin operator $s^+$ and the operator for the orbital angular momentum
$l^z$ commute, the following relation is satisfied,
\begin{eqnarray}
   s^+_{\tilde \alpha \tilde \beta} s^+_{\tilde \beta \tilde \alpha} &=&
        \langle \alpha | U^{-1} s^+ U | \beta \rangle 
        \langle \beta | U^{-1} s^+ U | \alpha \rangle
\nonumber \\ 
        &=&
        e^{-2 i\varphi} s^+_{\alpha \beta} s^+_{\beta \alpha} \, .
\end{eqnarray}
As a consequence, the partial summation in Eq.~(\ref{E++0}) over all 
wavevectors ${\bf q}$ which are equivalent due to symmetry yields a factor
$(1+e^{-i\pi}+e^{-2i\pi}+e^{-3i\pi})=0$ for the easy axis $\langle 100 \rangle$
and $(1+e^{-4i\pi/3}+e^{-8i\pi/3})=0$ for $\langle 111 \rangle$, i.e.,
$E^{++}_{k=0}=0$ in both cases.
In contrast, the corresponding factor $(1+e^{-2i\pi})$ for 
$\langle 110 \rangle$ is nonzero, and $E^{++}_{k=0}$ is, in general, finite.

\subsection{Relation between energy gap and $E_{k = 0}^{+-}$}

To determine the energy cost of tilting the spin polarization by small angle
$\theta$ out of the mean-field direction, we add to the Hamiltonian the 
perturbation
\begin{equation}
   H' = \Delta \left[ -{\theta^2\over 2}s^z + 
        \theta \left( s^x \cos \varphi + s^y \sin \varphi \right) \right] \, ,
\label{pert}
\end{equation}
and use standard perturbation theory.
Here, $\varphi\in [0,2\pi)$ is the azimuth, and $\Delta=N_{\rm Mn}J_{\rm pd}S$.
The linear order in $\theta$ does not contribute since 
$\langle s^x \rangle = \langle s^y \rangle = 0$.
To obtain the quadratic order in $\theta$ we use first-order perturbation 
theory for the $s^z$ term in Eq.~(\ref{pert}) and second-order perturbation 
theory for the $s^x \cos \varphi + s^y \sin \varphi = 
(s^+ e^{-i\varphi} + s^- e^{i\varphi})/2$ contribution.
For the former we get
\begin{equation}
   \delta E' = - {\theta^2 \Delta \over 2V} \sum_{\bf q} \sum_{\alpha}
        f [\epsilon_\alpha({\bf q})] \langle \alpha | s^z | \alpha \rangle 
        = {\theta^2 \Delta p\xi \over 4} \, ,
\end{equation}
and the result for latter reads
\begin{eqnarray}
   \delta E'' &=& {\theta^2 \Delta^2 \over 4V} \sum_{\bf q}
        \sum_{\alpha\beta} f [\epsilon_\alpha({\bf q})] 
        { | \langle \alpha | s^+ e^{-i\varphi} + s^- e^{i\varphi} | \beta 
        \rangle |^2 \over \epsilon_\alpha({\bf q}) - \epsilon_\beta ({\bf q})} 
\nonumber \\
   &=& -{\theta^2\Delta^2\over 8}\left( 2 E_{k = 0}^{+-} 
        + e^{-2i\varphi} E_{k = 0}^{++} + e^{2i\varphi} E_{k = 0}^{--} \right)
        \, .
\end{eqnarray}
If the easy axis is along $\langle 100 \rangle$ or $\langle 111 \rangle$, then
$E_{k = 0}^{++} = E_{k = 0}^{--} =0$, and the energy is independent of 
$\varphi$.
The spin wave energy at $k=0$ can now be obtained from the ratio of the 
energy change $\delta E'+\delta E''$ and the change of the spin 
$\delta S = \theta^2N_{\rm Mn}S/2$, which yields
\begin{equation}
  {\Omega_{k=0}\over \Delta} = { J_{\rm pd} \over 2}
        \left( { p\xi \over \Delta } - E_{k=0}^{+-} \right) \, ,
\end{equation}
and, as desired, we recover the $k=0$ limit of Eq.~(\ref{dispersion}) for an
easy-axis direction $\langle 100 \rangle$ or $\langle 111 \rangle$.

\end{appendix}

\end{document}